\documentclass[11pt, oneside]{article}   	
\usepackage{geometry}                		
\usepackage{graphicx}				
\usepackage{amssymb}

\usepackage{url}

\title{\bf Eigenvalues for a Pure Quartic Oscillator}
\author{S. M. Blinder\footnote{email: sblinder@wolfram.com} \\ Wolfram Research Inc., Champaign, IL 61820-7237 USA}
\date{}							

\begin{document}
\maketitle

\centerline{\Large\bf Abstract}
\vspace{.25cm}
\noindent The eigenvalues of a pure quartic oscillator are computed, applying a canonical operator formulation, generalized from the harmonic oscillator. Solving a 10x10 secular equation produces eigenvalues in agreement, to at least 4 significant figures, with accurate computations given in the literature.

The oscillator with a quartic anharmonicity, with Hamiltonian
\begin{equation}
H=\frac{p^2}{2\mu}+\frac{1}{2} k x^2+\frac{1}{4} \lambda x^4
\end{equation}
has been extensively treated in the literature [1,2]. This note will consider the pure quartic oscillator, in which the quadratic term is missing: $k=0$. For simplicity, we take $\mu=1, \hbar=1$. The Schr\"odinger equation we consider thus reduces to
\begin{equation}
-\frac{1}{2}\psi''(x)+\frac{1}{4} \lambda x^4\psi(x) = E \psi(x).
\end{equation}
The coordinate substitution $x\to \lambda^{-1/6} X$ transforms the equation into
\begin{equation}
-\frac{1}{2}\Psi''(X)+\frac{1}{4} x^4\Psi(X) = E \lambda^{-1/3}  \Psi(X).
\end{equation}
Thus only the case $\lambda=1$ need be considered, with the energy scaling as $\lambda^{1/3}$ for $\lambda \neq 1$. No analytic solution of this problem has been found, but a number of accurate numerical computations have been carried out [3,4].

It is useful for general orientation to compare the WKB computation of the eigenvalues. The requisite equation is
\begin{equation}
\sqrt{2}\oint \sqrt{E-\frac{\lambda}{4} x^4}\ dx= 2 \sqrt{2} \int_{-(4E/\lambda)^\frac{1}{4}}^{(4E/\lambda)^\frac{1}{4}}
\sqrt{E-\frac{\lambda}{4} x^4}\ dx= 
\frac{16}{3} K(-1) \frac{E^{3/4}}{\lambda^{1/4}}=\left(n+\frac{1}{2}\right)2\pi,
\end{equation}
Where $K$ is a complete elliptic integral of the first kind.
This givies the energies
\begin{equation}
E_n = \frac{3^{4/3}\pi^2} {2^{2/3} \Gamma(\frac{1}{4})^{8/3}}\left(n + \frac {1} {2}\right)^{4/3}\lambda^{1/3}=0.867145\left(n + \frac {1} {2}\right)^{4/3}\lambda^{1/3}, \ 
n=0, 1, 2, \dots.
\end{equation}
The first 10 WKB energies, for $\lambda=1$, are tabulated below:
\begin{eqnarray*}
E_0=0.344127,  \ E_1=1.48895,  \ E_2=2.94224, \ E_3=4.60804,  \ E_4=6.44231, \\
E_5=8.41864,  \ E_6=10.519,  \ E_7=12.7303,  \ E_8=15.0424,  \ E_9=17.4471.
\end{eqnarray*}

We can
 obtain an accurate approximation to the eigenvalues of the quartic oscillator by an operator method, previously applied to the inversion of ammonia [5].
Since the Hamiltonian contains only even powers of $p$ and $x$, a representation based on the ladder operators $a$ and $a^\dagger$ suggests itself, a generalization of the canonical operator formulation for the harmonic oscillator. Accordingly, we define
\begin{equation}
a = \sqrt{\frac{\omega}{2}}\, x + i\sqrt{\frac{1}{2 \omega}}\, p, \qquad
a^\dagger = 
 \sqrt{\frac{\omega}{2}}\, x - i \sqrt{\frac{1}{2  \omega}} \,p.
\end{equation}
The parameter $\omega$ is introduced, with its value to be determined such as to optimize the results.
The actions of the ladder operators on a basis ket are given by
\begin{equation} 
a|n\rangle = \sqrt{n}|n-1\rangle,  \qquad a^\dagger |n\rangle = \sqrt{n+1}|n+1\rangle.
\end{equation}
so that
\begin{equation}
x |n\rangle =  \sqrt{\frac{1}{2 \omega}} \Big(\sqrt{n}|n-1\rangle + \sqrt{n+1}|n+1\rangle\Big)
\end{equation}
and
\begin{equation}
p |n\rangle= -i \sqrt{\frac{ \omega}{2}} \Big(\sqrt{n}|n-1\rangle - \sqrt{n+1}|n+1\rangle\Big).
\end{equation}
By successive application of these operators, it follows that
\begin{equation}
x^2 |n\rangle =  \frac{1}{2  \omega} \Big(\sqrt{n(n-1)} |n-2\rangle +(2n+1) |n\rangle +\sqrt{(n+1)(n+2)} |n+2\rangle \Big)
\end{equation}
and
\begin{equation}
p^2 |n\rangle =  -\frac{ \omega}{2} \Big(\sqrt{n(n-1)} |n-2\rangle -(2n+1) |n\rangle +\sqrt{(n+1)(n+2)} |n+2\rangle \Big).
\end{equation}
Note, incidentally, that
\begin{equation}
\Big(\frac{p^2}{2}+\frac{1}{2}\omega^2 x^2 \Big) |n\rangle =\left(n+\frac{1}{2}\right)\omega |n\rangle,
\end{equation}
which agrees with the result for an harmonic oscillator. Finally, we require
\begin{eqnarray}
x^4 |n\rangle =  \frac{1}{4 \omega^2} \Big(\sqrt{n(n-1)(n-2)(n-3)} |n-4\rangle+
2\sqrt{n(n-1)}(2n-1) |n-2\rangle+ \hspace{1cm} \nonumber \\
(6n^2+6n+3)  |n\rangle+ 
2\sqrt{(n+1)(n+2)} (2n+3)  |n+2\rangle+ \hspace{1cm} \nonumber \\
\sqrt{(n+1)(n+2)(n+3)(n+4)} |n+4\rangle \Big). \hspace{1cm}
\end{eqnarray}

The nonzero matrix elements  of the Hamiltonian are given by
\begin{equation}
H_{n,n}= \frac {\omega} {4}(2n+1)+  \frac{1}{16\omega^2}(6 n^2 + 6 n + 3),
\end{equation}
\begin{equation}
H_{n+2,n}=H_{n,n+2}= -\frac {\omega}{4}\sqrt {(n + 1) (n + 2)}+\frac{1}{8\omega^2}\sqrt {(n + 1) (n + 2)}
 (2 n + 3),
\end{equation}
\begin{equation}
H_{n+4,n}=H_{n,n+4}=  \frac{1}{16\omega^2} \sqrt{(n+1)(n+2)(n+3)(n+4)}.
\end{equation}

Good convergence is obtained with the value $\omega=2.16$. The 10$\times$10 truncated matrix of $H$ is shown below:
\begin{figure}[h]
\begin{center}
\includegraphics[height=3cm]{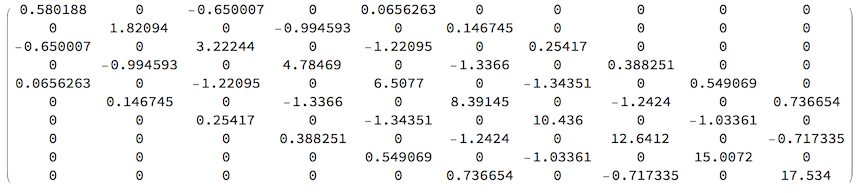}
\caption{Matrix $H_{mn}$.}
\label{mx}
\end{center}
\end{figure}

\noindent The eigenvalues are calculated using the  \url{Eigenvalues} program in 
{\it Mathematica}\textsuperscript{TM}.
The lowest 10 eigenvalues are tabulated below:
\begin{eqnarray*}
E_0=0.420805, \ E_1=1.5079, \ E_2=2.95886, \ E_3=4.62127,\ E_4= 6.46063,   \nonumber \\  
E_5=8.43686, \ E_6=10.6016, \ E_7=12.876, \ E_8=15.3116, \ E_9=17.7303.
\end{eqnarray*}
Our computations agree with the published results for the quartic oscillator to at least 4 significant figures.

Below is a plot of the potential energy of a quartic oscillator $V(x)=\frac{1}{4}x^4$, on which is superposed the computed energies $E_0, E_1, \dots, E_9$, as horizontal red lines. For comparison the corresponding WKB energies are also shown as gray lines.
\begin{figure}[h]
\begin{center}
\includegraphics[height=10cm]{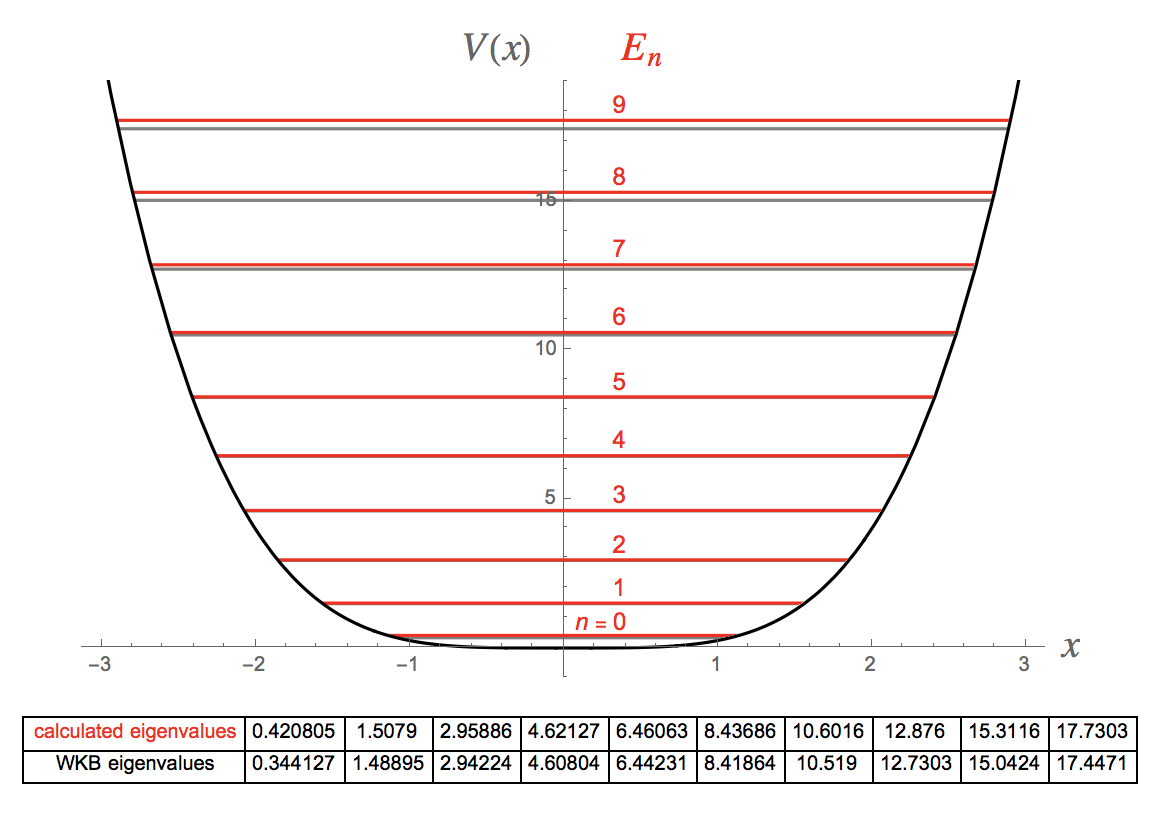}
\caption{Plot of $V(x)$, showing calculated eigenvalues as red lines, WKB energies as gray lines.}
\label{plot}
\end{center}
\end{figure}

\vspace{10cm}
\leftline{\Large\bf References} 

\noindent A preliminary version of this computation: 
S. M. Blinder "Eigenvalues for a Pure Quartic Oscillator"
\url{http://demonstrations.wolfram.com/EigenvaluesForAPureQuarticOscillator/}
 Wolfram Demonstrations Project
Published: March 15, 2019.

\noindent [1] F. T. Hioe and E. W. Montroll, "Quantum theory of anharmonic oscillators. I. Energy levels of oscillators with positive quartic anharmonicity," {\it Journal of Mathematical Physics}, {\bf 16}, 1945 (1975).

\noindent [2] S. Mandal, "Quantum oscillator of quartic anharmonicity,"
{\it Journal of Physics A: Mathematical and General}, {\bf 31}, L501 (1998).

\noindent [3] S. N. Biswas, K. Datta, R. P. Saxena, P. K. Srivastava and V. S. Varma,
"Eigenvalues of $\lambda \,x^{2 m}$ anharmonic oscillators,"
{\it Journal of Mathematical Physics} {\bf 14}, 1190 (1973).

\noindent [4] P. M. Mathews, M. Seetharaman, S. Rraghavan and V. T. A. Bhargava, "A Simple Accurate Formula for the Energy Levels of Oscillators with a Quartic Potential," {\it Physics Letters} {\bf 83A}(3), 118 (1981).

\noindent [5] S. M. Blinder, Ammonia Inversion Energy Levels using Operator Algebra, \url{arXiv:1809.08178} (2018).

\end{document}